%% file: ICPR_2020_v1.tex
\documentclass[10pt,a4paper,conference]{IEEEtran}

\usepackage{amsmath,amssymb,mathtools}
\usepackage{stmaryrd}
\usepackage{enumerate}
\usepackage{array}
\usepackage{eqparbox}
\usepackage{frame}
\usepackage{float}
\usepackage[ruled, lined, linesnumbered, commentsnumbered, longend]{algorithm2e}
\usepackage{color}
\usepackage[tight,footnotesize]{subfigure}
\usepackage{url}
\usepackage{epsfig}
\usepackage{multirow}
\usepackage{color}
\usepackage{subfigure}
\usepackage{cite}
\usepackage{todonotes}
\usepackage[font=small]{caption}
\usepackage{balance}
\usepackage{graphicx}
\usepackage{graphbox}
\usepackage[export]{adjustbox}
\usepackage{booktabs}

\input{zcom}

\newcommand{\argmin}{\mathrm{argmin}}

\begin{document}
	\title{Learning to segment clustered amoeboid cells from brightfield microscopy via multi-task learning with adaptive weight selection}
	
	\author{\IEEEauthorblockN{Rituparna Sarkar, Suvadip Mukherjee, Elisabeth Labruy\`ere and  Jean-Christophe Olivo-Marin }
		\IEEEauthorblockA{Bioimage Analysis Unit \\
			Department of Cell Biology and Infection
			 Institut Pasteur, Paris, France\\
			\textit{Email: \{rituparna.sarkar, smukherj, elisabeth.labruyere, jcolivo\}@pasteur.fr}} %
			\thanks{This work was partially funded through grants from the Labex IBEID (ANR-10-LABX-62-IBEID), France-BioImaging infrastructure (ANR-10-INBS-04), and the programs PIA INCEPTION (ANR-16-CONV-0005) and DIM ELICIT Région Ile de France.}}

\maketitle
\begin{abstract}	
	Detecting and segmenting individual cells from microscopy images is critical to various life science applications. Traditional cell segmentation tools are often ill-suited for applications in brightfield microscopy due to poor contrast and intensity heterogeneity, and only a small subset are applicable to segment cells in a cluster. In this regard, we introduce a novel supervised technique for cell segmentation in a multi-task learning paradigm. A combination of a multi-task loss, based on the region and cell boundary detection, is employed for an improved prediction efficiency of the network. The learning problem is posed in a novel min-max framework which enables adaptive estimation of the hyper-parameters in an automatic fashion. The region and cell boundary predictions are combined via morphological operations and active contour model to segment individual cells.
	The proposed methodology is particularly suited to segment touching cells from brightfield microscopy images without manual interventions. Quantitatively, we observe an overall Dice score of $0.93$ on the validation set, which is an improvement of over $15.9\%$ on a recent unsupervised method, and outperforms the popular supervised U-net algorithm by at least $5.8\%$ on average.
	

\end{abstract}
\section{Introduction}	
\begin{figure}[t]
	\centering
	\renewcommand{\tabcolsep}{0.05cm}
	\begin{tabular}{@{}ccc@{}}		
		\includegraphics[width=0.30\linewidth,height=0.30\linewidth ]{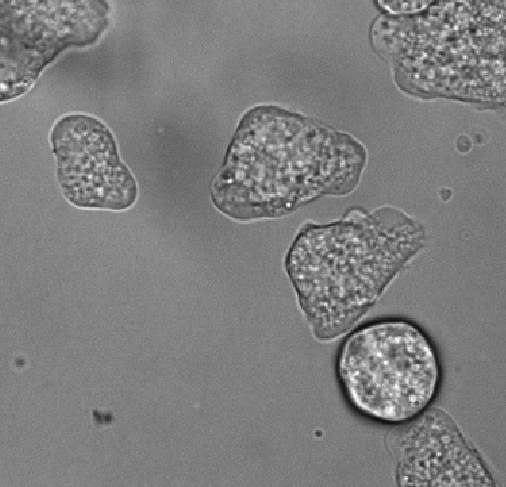}
	& \includegraphics[width=0.30\linewidth,height=0.30\linewidth]{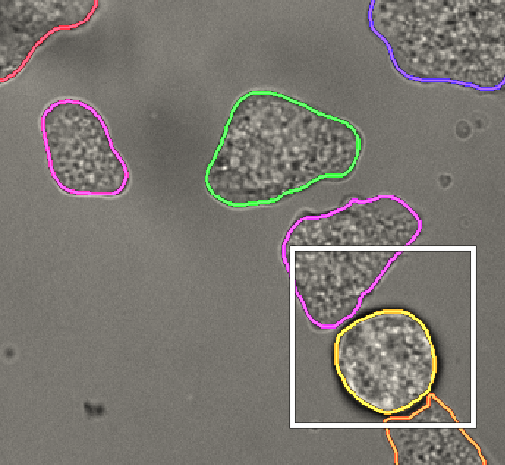}
		& \includegraphics[width=0.30\linewidth,height=0.30\linewidth]{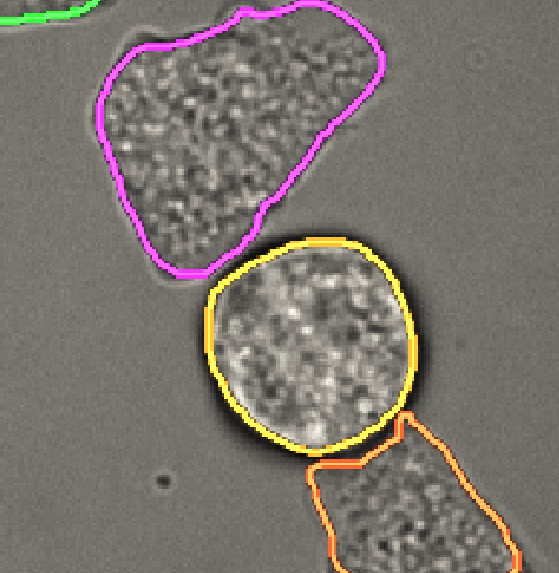}
		\\
		\includegraphics[width=0.30\linewidth,height=0.30\linewidth]{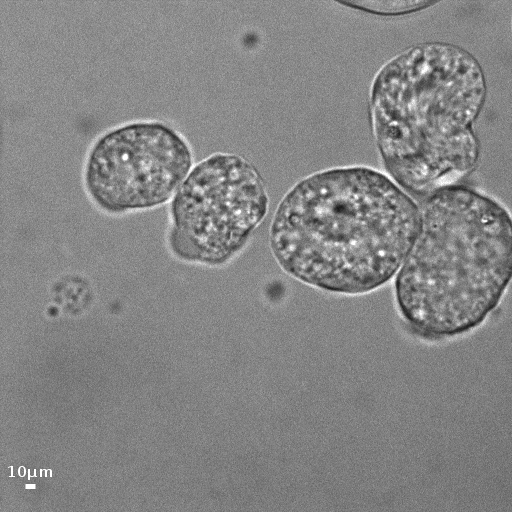}
		& \includegraphics[width=0.30\linewidth,height=0.30\linewidth]{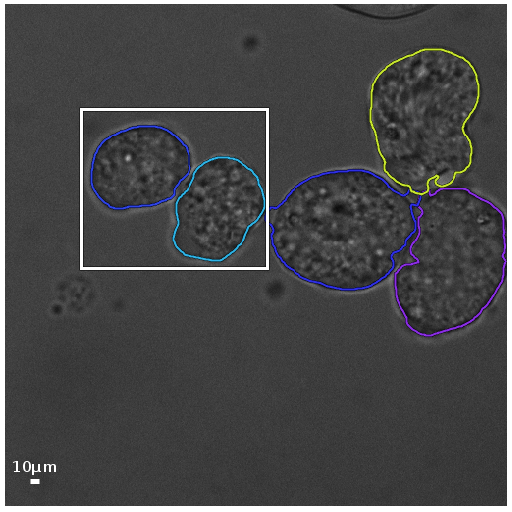}
		& \includegraphics[width=0.30\linewidth,height=0.30\linewidth]{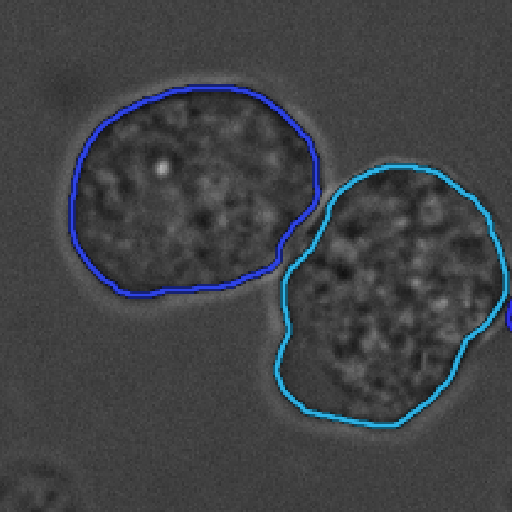}
	\end{tabular}
	\small \caption{Two representative brightfield images of cells are shown in the first column. The second column displays the segmentation results using our methodology, and the enclosed regions are zoomed in the last column to highlight the our method's ability to delineate touching cells. The images in the first column have been contrast stretched for intensity normalization. The images are best viewed in color.}
	\label{fig: motion_dir}
\end{figure}

Understanding the mechanisms of cell deformation and cell motility is critical to several unsolved problems in cell biology. The implications of cell motility in  biological processes such as embryonic development, cancer metastasis etc. \cite{tweedy2013distinct, dufour2015amoeboid, dufour2014signal} have motivated subsequent research to analyze the morphological and bio-physical properties of motile cells\cite{tweedy2013distinct, dufour2015amoeboid,boquet2017bioflow}. 
In this paper we pay attention to cells exhibiting \textit{amoeboid} motion, which is a particular model of cell motility \cite{dufour2015amoeboid}.		
Amoeboid motion is characterized by repetitive formation of small protrusions of the cell membrane which are better visible via \textit{brightfield} microscopy. Automating such large scale experimental analyses invariably demands robust strategies for cell segmentation.  
However, poor contrast and  signal variation in brightfield images pose significant challenges, which are accentuated when the cells are in close proximity.  
We address these issues, and propose a novel cell segmentation technique for brightfield imagery with special emphasis on clustered objects (see Fig.~\ref{fig: motion_dir}). 
The segmentation problem is posed as a multi-task optimization strategy using deep convolutional neural network architecture, with adaptive estimation of the task parameters. 

\subsection{Background and motivation}
Traditional  segmentation techniques based on deformable models\cite{zimmer2005coupled, li2007active, mukherjee2014region} generalize poorly to brightfield images due to low contrast and blurred cell boundaries. The more recent techniques which are capable of handling intensity inhomogeneity \cite{mukherjee2014region, sarkar2015dictionary}, are better suited for segmenting isolated objects, but are less effective for clustered cells. Other developments in this field include the work of \cite{pecreaux2006biophysical}, where the model is constrained by the biophysical properties of the cell membrane. Most such approaches rely significantly on accurate model initialization which is non-trivial in a fully automated setting.
A recent work\cite{ali2012automatic} advocates a multi-focus strategy for automated initialization, but this may not be practically feasible in a laboratory setup as motile cells often exhibit significant displacement between consecutive captures. 

\begin{figure*}[t]
	\centering	
	\includegraphics[width=0.85\linewidth]{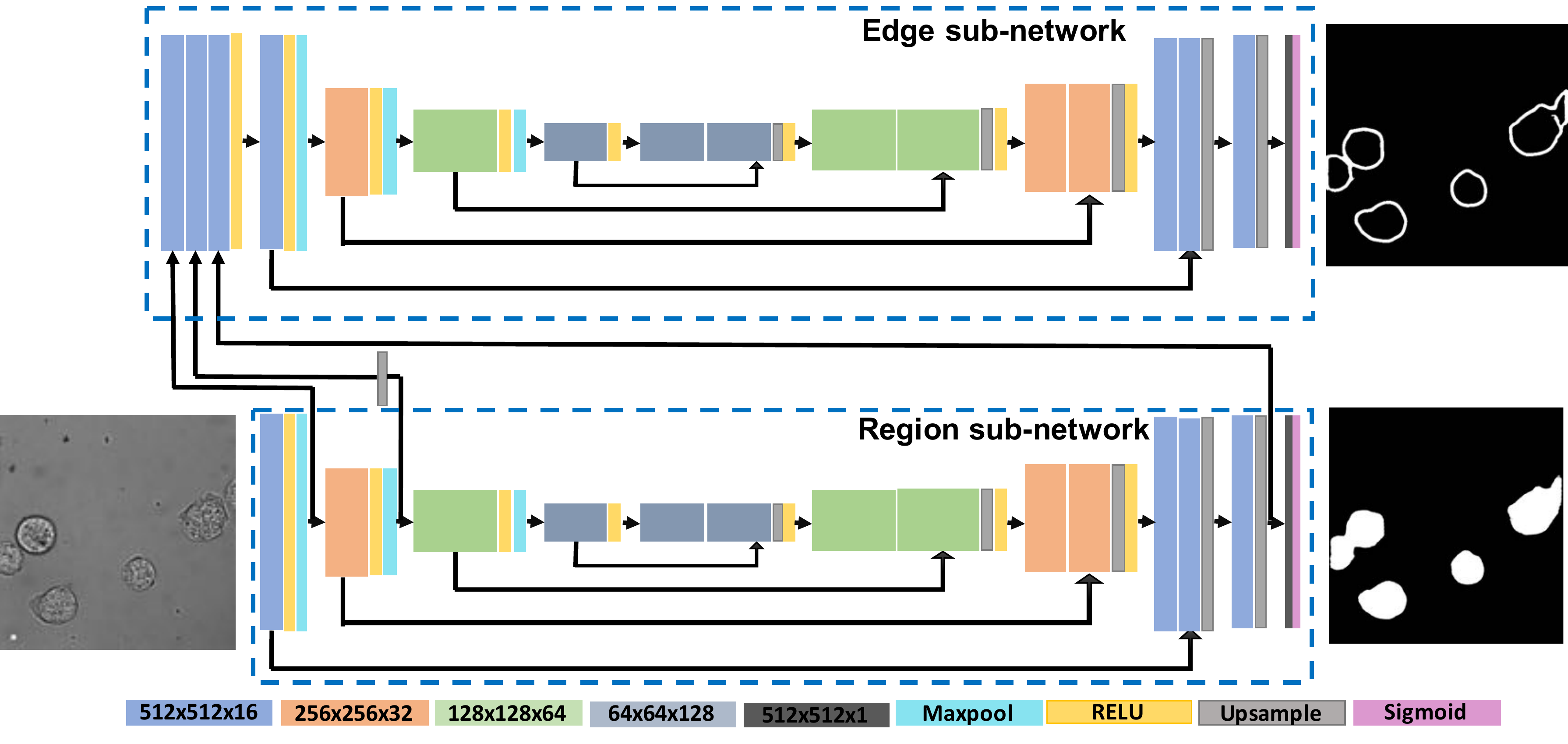}			
	\caption{Multi-AmoebaNet architecture.}
	\label{fig: multi-amoebanet}
\end{figure*} 

In the recent years deep convolutional neural networks (and their variants) have emerged as the \textit{de facto} standard for segmentation {\cite{ronneberger2015u, singhal2017automated, guerrero2018multiclass, schmidt2018cell, chen2016semantic,kendall2018multi}}. In particular, the U-net architecture \cite{ronneberger2015u} has shown promise in bio-imaging applications where sufficient (annotated) training samples may not be readily available. However, its efficacy to identify clustered cells is proportional to the number of such examples encountered during model training, which is a difficult constraint to satisfy in practice. U-net type networks are also known to be susceptible to signal in-homogeneity and poor  contrast\cite{ravishankar2017learning}, which hinder its use for brightfield images as a plug-and-play tool. 
Hybrid-learning \cite{singhal2017automated,mukherjee2017lung}, which combines deep neural network architectures with model based segmentation methods (such as level sets or graph cut) have shown promise for such complex tasks, although they are not particularly suited for continuous model update via end-to-end learning. 

Recently, the authors in \cite{ kendall2018multi, chen2016semantic} argue that the performance of supervised deep neural networks can be significantly improved by training them on multiple tasks simultaneously, as opposed to learning individual models for each task. The key principle is to oblige the system to learn intricate contextual cues by optimizing simultaneously for multiple task categories (such as object segmentation, depth prediction, object instance detection etc.) in an unified, end-to-end framework for better accuracy and reduced computational cost. Such multi-task learning strategies \cite{schmidt2018cell,guerrero2018multiclass, guerrero2019weakly} have been adopted for bio-imaging applications to tackle the specific challenges, such as, significant morphological variations of the cells, inhomogeneous signal intensity and scarcity of training data. An important design challenge for such techniques is to estimate the optimal combination of the penalty function for each task \cite{kendall2018multi}.

\subsection{Our contribution}
A majority of the aforementioned approaches use parameters which are either chosen na\"ively, or tuned manually, thereby introducing selection bias in the system. To address this issue, we formulate the multi-task problem as a min-max optimization, where the individual task parameters are estimated analytically at each step of the global model update.
The key contributions of this work are highlighted as follows:
\begin{itemize}
\item Novel convolutional neural network based  architecture for cell segmentation embedded in a multi-task framework with adaptive estimation of task weights;
\item Robust solution to reconstruct individual cell morphology when the objects are significantly clustered;
\item Significant improvement in performance over the state-of-the-art for segmenting (possibly) touching cells in brightfield microscopy.
\end{itemize}
The proposed methodology is presented next, followed by experimental evaluations and the concluding remarks in  Sec.~\ref{sec:expt} and Sec.~\ref{sec:conclusion} respectively.

\section{Method}
\begin{figure*}[t]
	\centering
	\renewcommand{\tabcolsep}{0.05cm}
	\begin{tabular}{@{}cccccc@{}}		
			\includegraphics[width=0.16\linewidth]{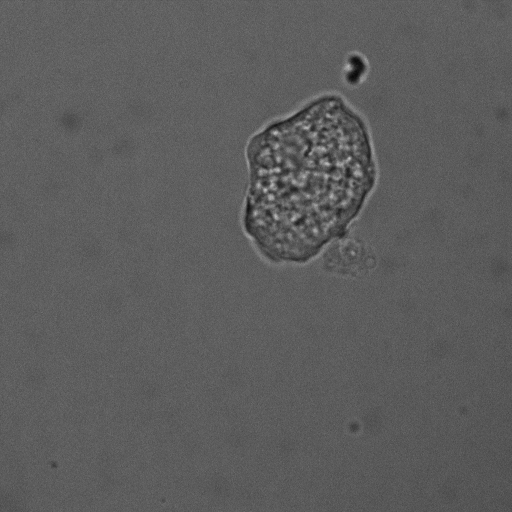}
			&\includegraphics[width=0.16\linewidth]{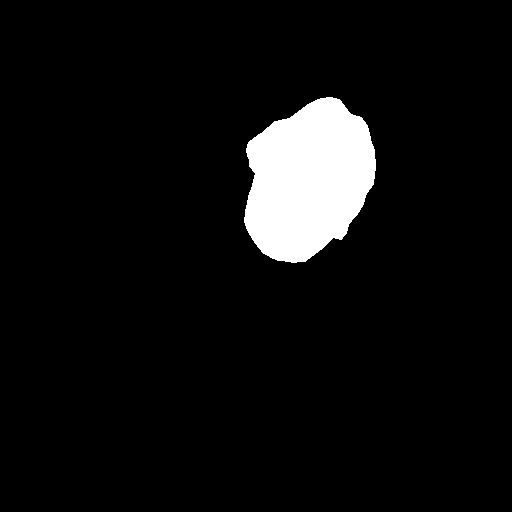}
		&		\includegraphics[width=0.16\linewidth]{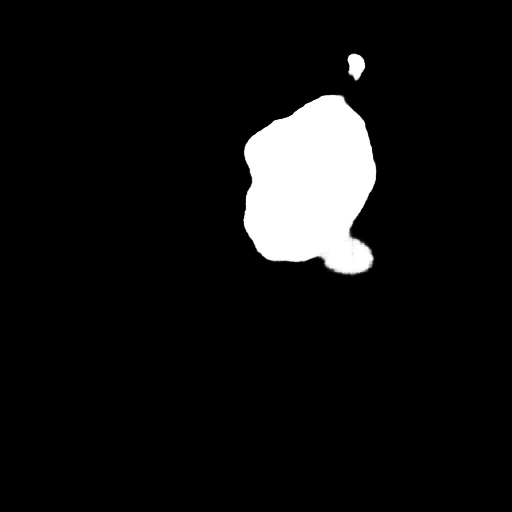}
				& \includegraphics[width=0.16\linewidth]{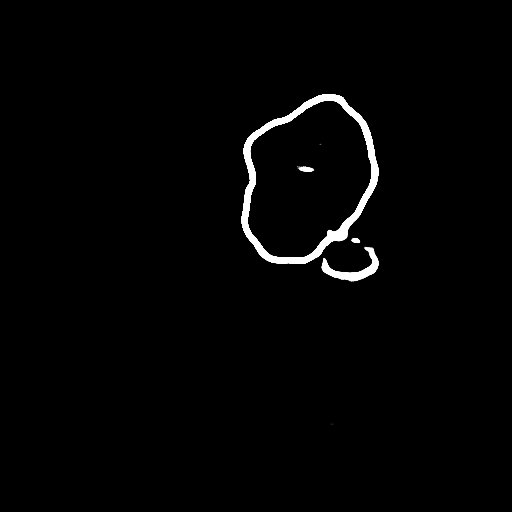}
				& \includegraphics[width=0.16\linewidth]{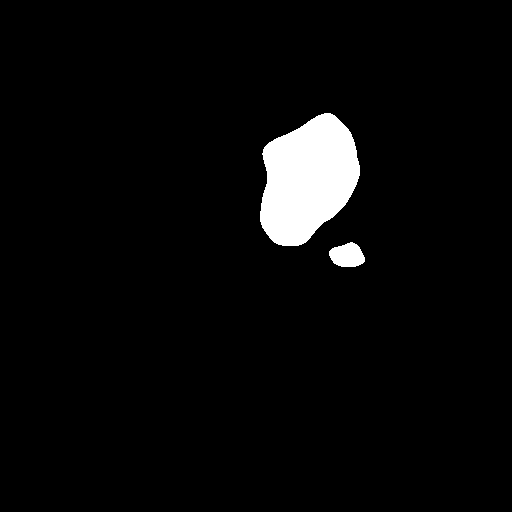}
				& \includegraphics[width=0.16\linewidth]{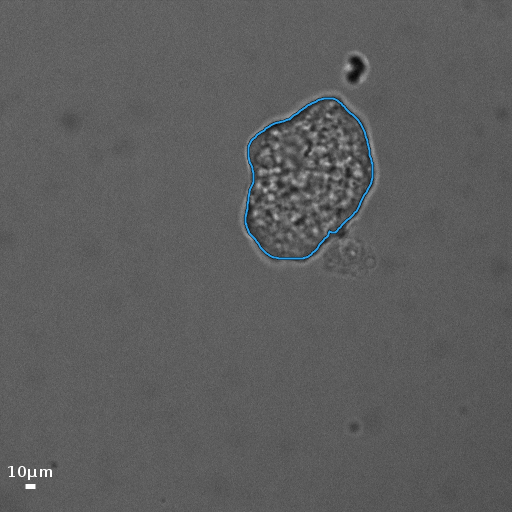} \\
		\includegraphics[width=0.16\linewidth]{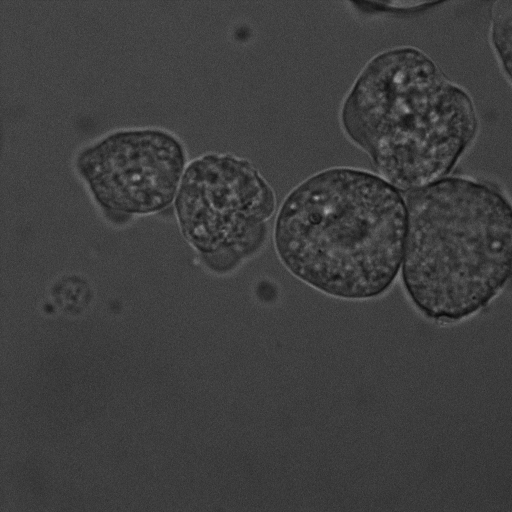} &
		\includegraphics[width=0.16\linewidth]{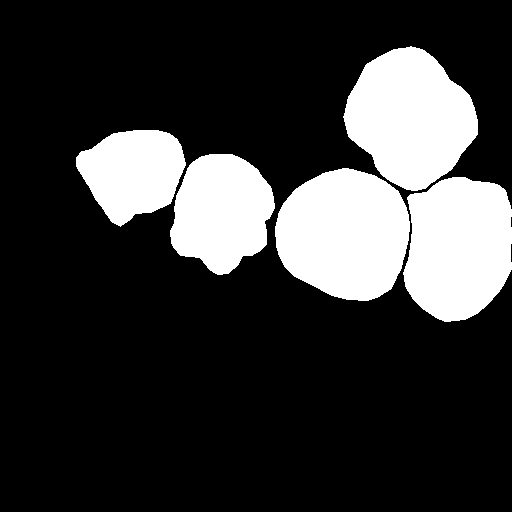}
		&\includegraphics[width=0.16\linewidth]{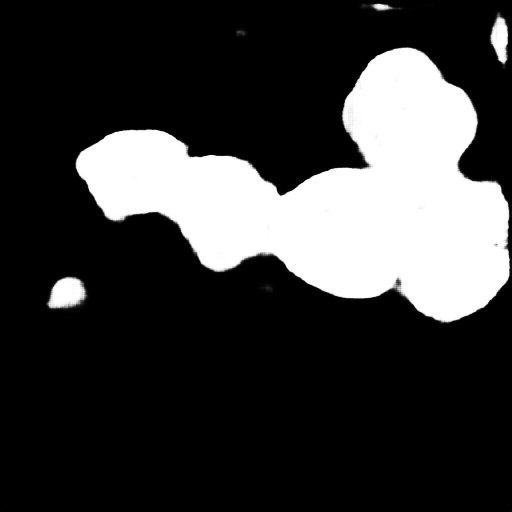}
		& \includegraphics[width=0.16\linewidth]{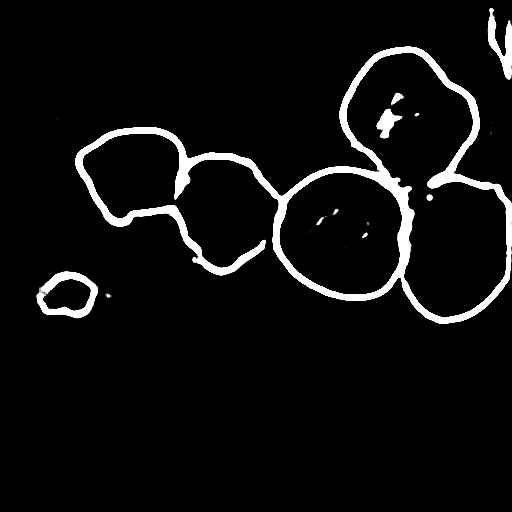}
		& \includegraphics[width=0.16\linewidth]{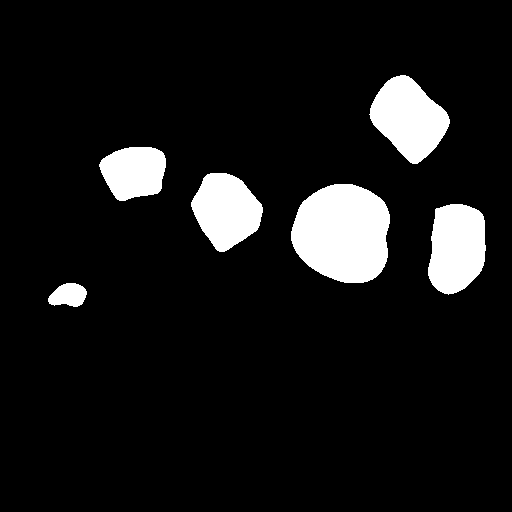}
		& \includegraphics[width=0.16\linewidth]{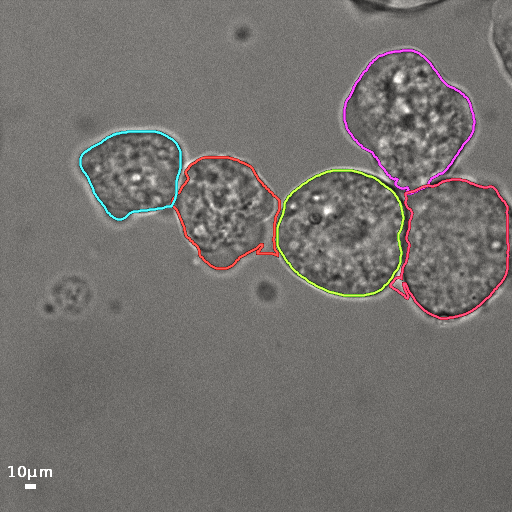}
		\\ \text(a) Original &   \text{(b) Ground Truth} &  \text{(c) Region} & \text{(d)\,Boundary}	& \text{(e)\, Initialization}&    \text{(f) Final Result}		
	\end{tabular}
	\caption{The original image, ground truth, region and edge predictions are shown in (a), (b), (c) and (d) respectively. The initial contour positions estimated via dynamic clustering are shown in (c), and the final segmentation results are shown in (d).}
	\label{fig: postprocessing}
\end{figure*}

In this section we formally introduce the proposed multi-task learning based framework for segmenting motile cells from brightfield microscopy. 
Accurate delineation of individual cells rely on precise estimation of cell location. Region based regression networks are often incapable to separate touching cells due to poor contrast at cell boundaries. Edge localization is essential to determine cell boundary, although when trained separately, such models often suffer from incomplete cell contours which leads to segmentation error due to leakage\cite{mukherjee2014region}. In summary, neither region prediction, nor edge indicators are adequately sufficient to segment cells from brightfield imagery (see Fig.~\ref{fig: postprocessing}). 
We therefore seek an optimal combination of the region and edge predictions to precisely segment cells from a dense cluster.
The proposed multi-task convolutional neural network (CNN) architecture, namely \textit{Multi-CellNet}, is trained to simultaneously predict the cell foreground and the cell boundary with adaptive estimation of the task hyper-parameters. The network architecture, and the optimization details are presented in the following subsections.

\subsection{Multi-CellNet architecture}

The designed network, shown in Fig.~\ref{fig: multi-amoebanet} consists for two sub-networks, viz. a \textit{region} and an \textit{edge} sub-network which will provide  pixel wise predictions of cell foreground region and cell boundary respectively.
Each sub-network represents an U-net type architecture\cite{ronneberger2015u}. Each sub-network consist of four down-sampling and four up-sampling blocks with skip connections. The sub-networks are coupled via network weight sharing at multiple levels.
Relu activation is applied after each convolution block in the network. Sigmoid activation is applied to the final layer to map the image output between 0 and 1. The detailed architecture is shown in Fig.~\ref{fig: multi-amoebanet}.
The region sub-network is designed to predict the cell foreground localization by penalizing its output with respect to the manually annotated cell masks. The edge sub-network is trained to predict an edge function for individual cells, which is critical to identifying the boundary between touching objects.
The gradient magnitude of the region masks smoothed via a Gaussian filter is used for supervision of the edge sub-network. The feature maps from the first two down-sampling and last up-sampling block of the region sub-network are concatenated and input to the edge sub-network. This ensures an unified training of the overall architecture and  contextual information sharing between the sub-networks.
Weighted combination of Dice loss for both sub-networks is used to train the network. 
In order to obtain an optimal combination of the loss, we estimate the loss weights using minmax optimization framework as described next.

\subsection{Model training with minimax optimization}
We represent an image defined on the domain $\Omega\subset\mathbb{R}^2$ as the function $f:\Omega\mapsto\mathbb{R}$ . The region and edge functions predicted by the sub-networks are represented by $f_r(x,y) = \mathcal{R}_{\theta_1}\left[f(x,y)\right]$ and $f_e(x,y) = \mathcal{H}_{\theta_2}\left[f(x,y)\right]$. Here, $\mathcal{R}_{\theta_1}$ and $\mathcal{H}_{\theta_2}$ are the functional approximates for the region and edge sub-networks respectively (with parameters $\theta_1$ and $\theta_2$), and both prediction outputs are normalized to $\left[0,1\right]$. Note that due to parameter sharing of our neural network architecture, $\theta_1\cap\theta_2\neq\emptyset$.
The energy function $\mathcal{E}$ for a multi-task network optimization is written as :
\begin{align}
\mathcal{E} & = \alpha E_1(\theta_1) + \beta E_2(\theta_2) 
\label{eq:optimization_sum}
\end{align}
Where $ \alpha$ and  $\beta$  are non negative parameters. A typical practice is to assign them equal values \cite{guerrero2019weakly, schmidt2018cell} or select them arbitrarily. However, a heuristic parameter selection strategy  could lead to imbalance in the contribution of the associated loss functions, resulting in a trained model which is biased towards a particular task. 
We propose an automated estimation of the task weights which are constrained as $\alpha^2 + \beta^2 = 1$.
Using this constraint, the network penalty function $\mathcal{E}$ can be defined as a nonlinear combination of the losses as follows:
\begin{align}
\mathcal{E}\left(\lambda, \theta_1, \theta_2\right) = \lambda E_1(\theta_1) + \sqrt{1-\lambda^2}E_2(\theta_2)
\label{eq:concave_sum}
\end{align}
Here, $\alpha = \lambda$ and $\beta = \sqrt{1-\lambda^2}$.
The loss functions $E_1$ and  $E_2$ are defined as, 
\begin{align}
E_1(\theta_1) =& 1-\mathcal{D}(g(f),f_r) \\ 
E_2(\theta_2) =& 1-\mathcal{D}(|\nabla g(f)|,f_e)
\label{eq:loss_unc} 
\end{align}
Here, $g(f)$ represents the ground truth annotation of the image $f$, and $|\nabla g(f)|$ is its Gaussian smoothed gradient magnitude. The function $\mathcal{D}$ is the regularized and differentiable Dice penalty loss \cite{ravishankar2017learning} which is computed as follows:
\begin{align}
\mathcal{D}(y, \hat{y}) = \frac{2\sum_{i=1}^N y_i  \hat{y}_i}{\sum_{i=1}^N y_i + \sum_{i=1}^N \hat{y}_i}
\end{align}
Here $y$ and $\hat{y}$ are two $N$-length vectors which typically correspond to the data label, and the network's prediction respectively. The combined loss for the multi-task network is computed via the non-linear weighting of $E_1$ and $E_2$. An analysis of Eq.~\ref{eq:concave_sum} reveals that $\mathcal{E}(.)$ is concave in $\lambda$. This follows directly by computing the partial derivatives as,
\begin{align}
\dfrac{\partial}{\partial\lambda}\mathcal{E}(\lambda, \theta_1, \theta_2) &= E_1 - \lambda\left(1-\lambda^2\right)^{-\dfrac{1}{2}}E_2  \label{eq:first_deriv} \\
\text{and}\quad \dfrac{\partial^2}{\partial\lambda^2}\mathcal{E}(\lambda, \theta_1, \theta_2) & = -E_2\left(1-\lambda^2\right)^{-\dfrac{3}{2}} \label{eq:sec_deriv}
\end{align}
Clearly, the second derivative in Eq.~\ref{eq:sec_deriv} is negative, since $E_2\in\left[0,1\right]$. The non-linear concave combination of the loss terms leads to a min-max optimization problem defined as 
\begin{align}
\min_{\theta_1,\theta_2}\max_\lambda \mathcal{E}(\lambda, \theta_1, \theta_2) \label{eq:minimax}
\end{align}
Intuitively, this formulation can indeed be interpreted as an optimization problem which seeks to minimize the worst case performance of the system. A benefit of this strategy is that the relative weights between the two different tasks can be obtained in an adaptive fashion using alternating minimization\cite{saha2009image}. The system parameters and the weight term $\lambda$ are obtained automatically via the following iterative scheme.
First, we compute the weight term which maximizes the total loss $\mathcal{E}$ by setting Eq.~\ref{eq:first_deriv} to zero which yields the following expression:
\begin{align}
\lambda^* = \frac{E_1(\hat{\theta_1})}{\sqrt{E_1(\hat{\theta_1})^2 + E_2(\hat{\theta_2})^2}}\label{eq:lambda}
\end{align}
The network parameters are then updated using stochastic gradient descent by solving 
\begin{align}
\hat{\theta_1}, \hat{\theta_2} = \underset{\theta_1,\theta_2}{\argmin}\,\mathcal{E}(\lambda^*, \theta_1, \theta_2), 
\end{align}
The above steps are iterated until convergence (see Algorithm 1). In practice, the Adam \cite{kingma2014adam} optimizer is used to estimate the network parameters, and the weight parameter is estimated via Eq.~\ref{eq:lambda} for each training epoch.

\subsection{Segmenting individual cells}

The region prediction from the \textit{Multi-CellNet} may not always be able to detect the cells separation, when cells are in close proximity (Fig.~\ref{fig: postprocessing}(\textit{a}). Similarly, the cell boundary can be mis-predicted when the cell and background contrast is low,  (Fig.~\ref{fig: postprocessing}(\textit{b})). As a result, appropriately combining the region and boundary predictions is necessary for smoother and accurate segmentation. The final cell separation and smooth contours are obtained by appropriately combining the cell region and boundary predictions embedded in an active contour model. 
Coupled active contours \cite{zimmer2005coupled,de2012icy}, which are specifically designed to prevent merging of adjacent parametric curves, is applied to obtain smooth boundaries of individual cells. 

First, automatic contour initialization if performed by estimating the initial cell localization using the variational, hierarchical clustering scheme due to \cite{yip2006dynamic}. This  ensures  the detection of  well-separated cellular regions (see Fig.~\ref{fig: postprocessing}\textit{c}) for active contour initialization. 
The initialized curves are then evolved outwards with the balloon force function $F_s = f_r(x,y)(1-f_e(x,y))$. The force function encourages curve motion for pixels with high region prediction score and low edge prediction value. The force function, further, compensates the false positives and false negatives in region prediction using the edge prediction and vice versa. Additionally, to prevent contour leakage due to inconsistent edge predictions, the model is implemented using the coupled active contour\cite{zimmer2005coupled} available as a plugin in the open source bioimage analysis toolkit, Icy \cite{de2012icy}. Convergence is achieved when the speed function restricts further curve propagation, or when the disjoint curves meet. The steps of the segmentation procedure are illustrated in Fig.~\ref{fig: postprocessing}.

\begin{algorithm}
\caption{Multi-CellNet Algorithm}
	\SetKwInOut{KwIn}{Input}
	\SetKwInOut{KwOut}{Output}
	\KwIn{Image: $f$; Region annotation: $g(f)$ ; Edge function: $|\nabla g(f)|$, Number of Epochs, $n$: 8000; Mini-Batch size, $m$: 10, Learning rate, $l_r$: $10^{-4}$ }
	\KwOut{Region prediction: $f_r(x,y)=\mathcal{R}_{\theta_1}\left[f(x,y)\right]$;  Edge prediction: $f_e(x,y) = \mathcal{H}_{\theta_2}\left[f(x,y)\right]$}
	Set number of epochs $n$. \\
	Initialize Multi-CellNet parameters $\theta_1$ and $\theta_2$\\
	\While{Number of Epochs less than $n$} {
	Obtain mini-batch: $\{f_1,|g(f_1)| ,|\nabla g(f_1)| \}$ $\dots$ $\{f_m, |g(f_m)|, \nabla g(f_m)|\}$ \\
	Compute loss over mini-batch: $E_1(\hat{\theta}_1) = \frac{1}{m}\sum_{j=1}^m {1-\mathcal{D}(g(f_j),f_r^{(j)}(x,y))}$ 
	 $E_2(\hat{\theta}_2) = \frac{1}{m}\sum_{j=1}^m{1-\mathcal{D}(|\nabla g(f_j)|,f_e^{(j)}(x,y))}$ \\
	Compute $\lambda^* = \frac{E_1(\hat{\theta_1})}{\sqrt{E_1(\hat{\theta_1})^2 + E_2(\hat{\theta_2})^2}}$ \\
	Compute gradients: $\partial \theta_1, \partial \theta_2 = \mathtt{Autograd}(E(\lambda^*, \theta_1, \theta_2))$ \\
	Update parameters: $(\hat{\theta}_1, \hat{\theta}_2) \leftarrow \mathtt{Adam}(\theta_1, \theta_2,\partial \theta_1,\partial \theta_2, l_r)$}
\end{algorithm}

\section{Experimental validation}
\label{sec:expt}
The Multi-CellNet is trained using 245 training images of size 512x512 aided with data augmentation (random combination of horizontal and vertical image flipping with linear and nonlinear intensity  scaling). The training is performed for epochs $n$ = 5000 and batch size, $m$ = 10. The learning rate is initialized to $10^{-4}$ with $0.99$ decay factor. The segmentation results are compared with the U-net \cite{ronneberger2015u} and L2S \cite{mukherjee2014region}, a deep-learning based and a classical image segmentation method respectively. The U-net model is trained using our in-house training data and binary cross-entropy loss. The loss function is weighted by the cell separation function as described by the original paper \cite{ronneberger2015u}. The U-net is trained for 8000 epochs with batch size of 10.  
L2S\cite{mukherjee2014region} is an region active contour based unsupervised segmentation technique which is initialized manually. 
In our experiments, we incorporate $3^{rd}$ order Legendre polynomial to model the image intensity.

\textbf{Quantitative Evaluation:} The Dice index and mean squared error is used to compare the quantitative performances of the different techniques. We evaluate a set of fifty test images consisting of single and multiple cells with varying magnification and illumination. The average scores and the standard deviation of the error are presented in Table \ref{table:dice}. We observe a Dice score of $0.77$ and $0.88$ for L2S and U-net respectively. On the same dataset, Multi-CellNet achieves a Dice index of $0.93$ which is an improvement of $16.4\%$ and $5.6\%$  over L2S and U-net respectively. Further, Multi-CellNet achieves a mean squared error of $0.11$ which is $4.4\%$ and $6.6\%$ improvement over U-net and L2S respectively. 
However, it may be noted that Dice coefficient as well as mean squared error computed over one image gives a notion of correct prediction of masks, but not cell separation. Hence, to evaluate efficacy of the methods in segmenting touching cells, we compute the Dice score and mean squared error for individual cells in the dataset. The error is computed between each cell in the ground truth image and its nearest neighbor in the predicted image. The average error over all the cells in the dataset and the standarad deviation of the error is presented in Table \ref{table:dicepercell}. As noted from the table, the Multi-AmoebaNet achieves an improvement of $4\%$ and $21\%$ over U-net \cite{ronneberger2015u} and L2S \cite{mukherjee2014region} respectively. We note that during the segmentation pipeline discussed in Section II.C, objects with significantly small size are discarded since they indicate either partial cells or extracellular objects. However, in some scenarios, partial cells are annotated in the ground truth. This leads to a significantly low (high) value in Dice score (mean squared error), and hence a decrease (increase) in the average error in comparison to the error computed for the entire image (Table \ref{table:dice}) is noticed.

\begin{center}
\begin{table}[t]
	\caption{Average error for the dataset.}
	\centering
	\begin{tabular}{ cccc } 
		\toprule
		\, & \textbf{U-net} \cite{ronneberger2015u} & \textbf{L2S} \cite{mukherjee2014region} & \textbf{Proposed}\\
		\midrule
		Dice & 0.88 $\pm$ 0.14 & 0.78 $\pm$ 0.23 & \textbf{0.94 $\pm$ 0.02} \\
		MSE & 0.15 $\pm$ 0.10 & 0.17 $\pm$ 0.10 & \textbf{0.11 $\pm$ 0.03} 	\\
		\bottomrule
	\end{tabular}
	\label{table:dice}
\end{table}
\end{center}

\begin{center}
\begin{table}[h]
	\caption{Average error for individual cells.}
	\centering
	\begin{tabular}{ cccc } 
		\toprule
		\, & \textbf{U-net} \cite{ronneberger2015u} & \textbf{L2S} \cite{mukherjee2014region} & \textbf{Proposed}\\
		\midrule
		Dice & 0.79 $\pm$ 0.30 & 0.62 $\pm$ 0.35 & \textbf{0.83 $\pm$ 0.27} \\
		MSE & 0.07 $\pm$ 0.07   & 0.12 $\pm$ 0.09 & \textbf{ 0.06$\pm$ 0.06 } 	\\
		\bottomrule
	\end{tabular}
	\label{table:dicepercell}
\end{table}
\end{center}

\begin{figure*}[t]
	\centering
	\renewcommand{\tabcolsep}{0.05cm}
	\begin{tabular}{@{}ccccccccc@{}}		
		\scriptsize{\rotatebox[origin=c]{90}{ Original}}&
		\includegraphics[valign=m,width=0.11\linewidth]{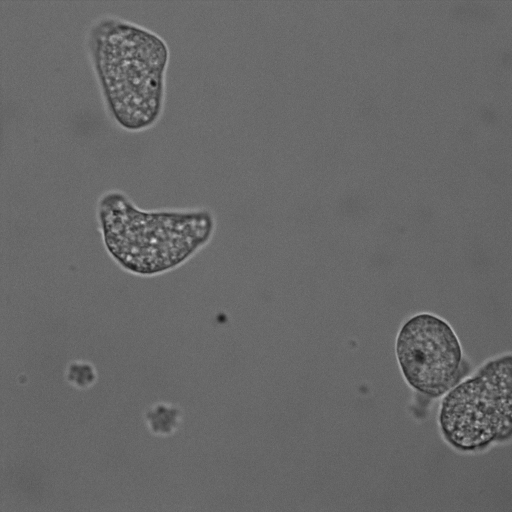}
		&\includegraphics[valign=m,width=0.11\linewidth]{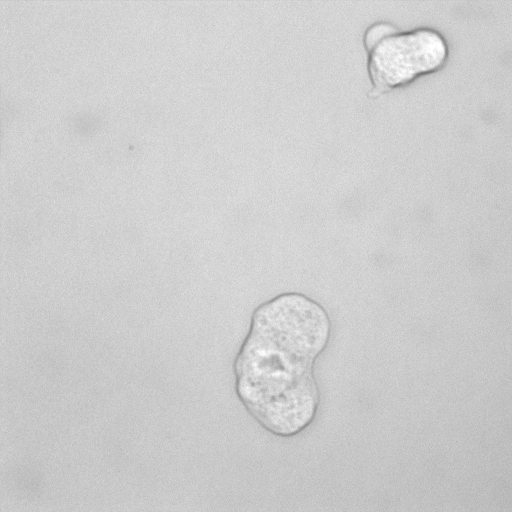}
		&\includegraphics[valign=m,width=0.11\linewidth]{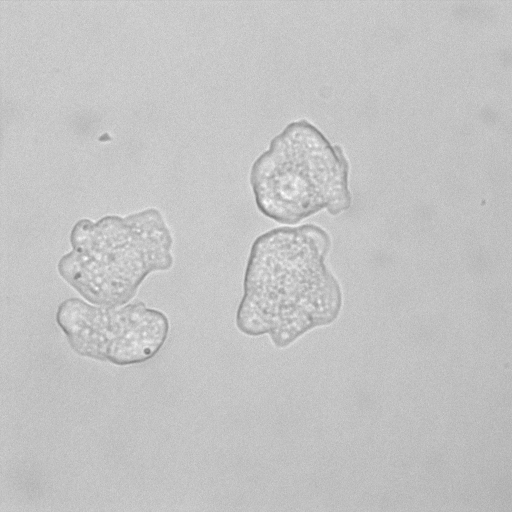}
		&\includegraphics[valign=m,width=0.11\linewidth]{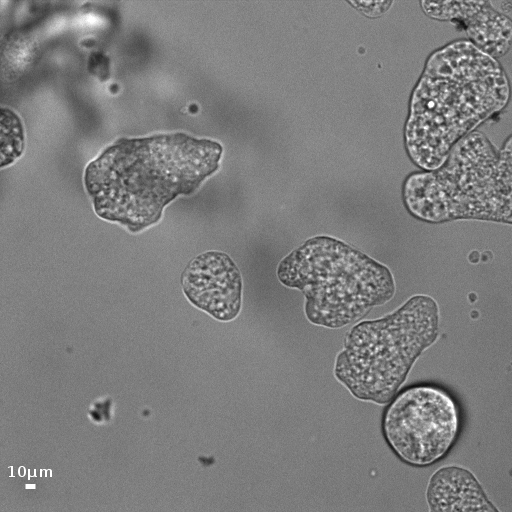}
		&\includegraphics[valign=m,width=0.11\linewidth]{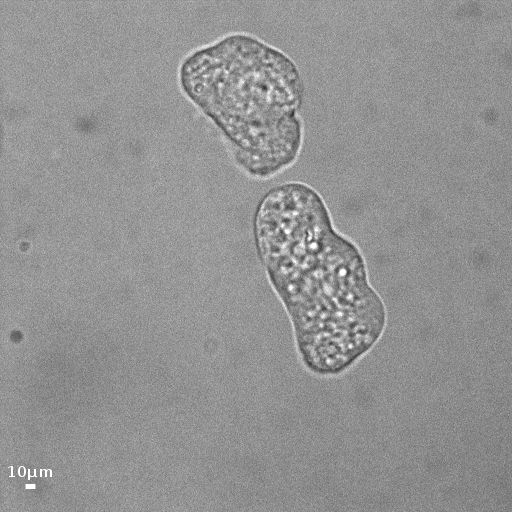}
		&\includegraphics[valign=m,width=0.11\linewidth]{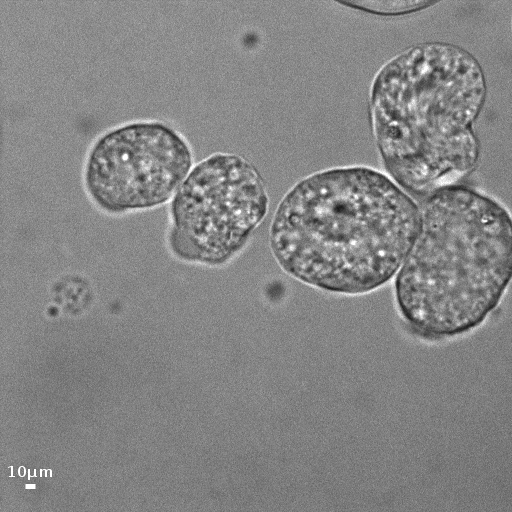}
		&\includegraphics[valign=m,width=0.11\linewidth]{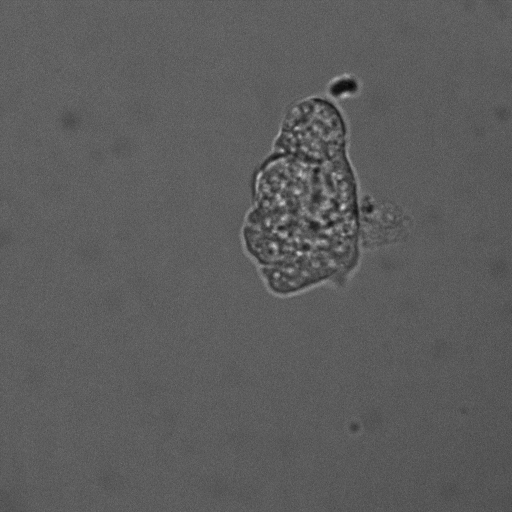}
		&\includegraphics[valign=m,width=0.11\linewidth]{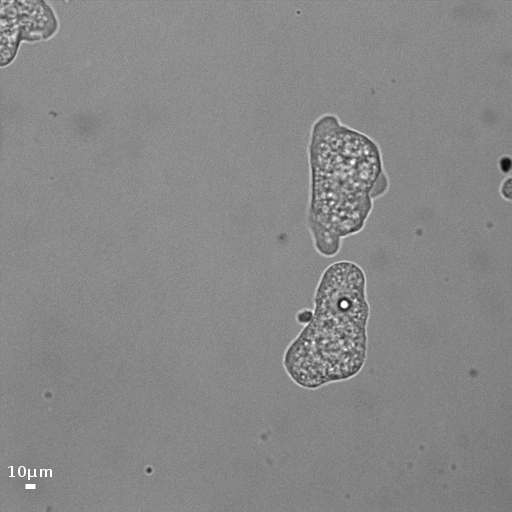}
		\\ 
		\scriptsize{\rotatebox[origin=B]{90}{Unet}} &
		\includegraphics[valign=m,width=0.11\linewidth]{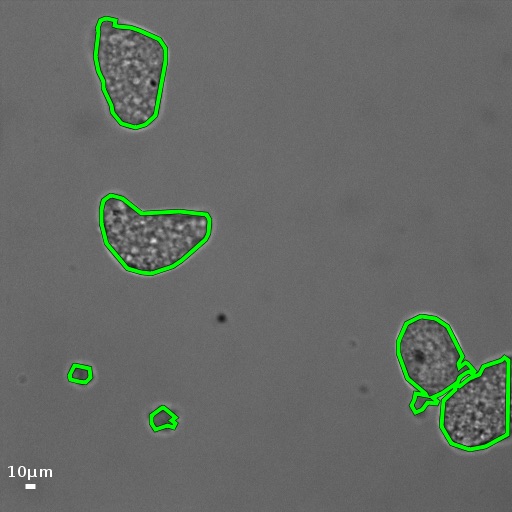}
		&\includegraphics[valign=m,width=0.11\linewidth]{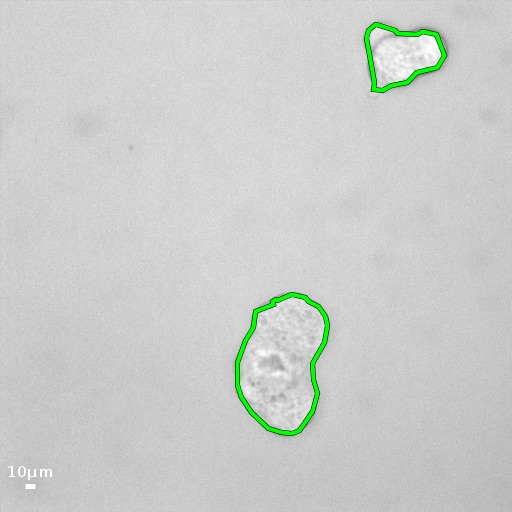}
		&\includegraphics[valign=m,width=0.11\linewidth]{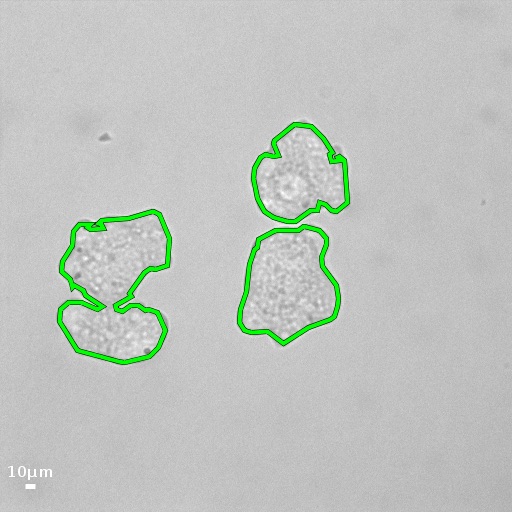}
		&\includegraphics[valign=m,width=0.11\linewidth]{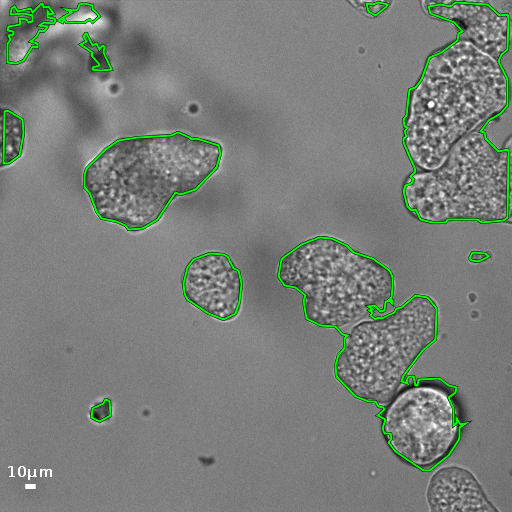}
		&\includegraphics[valign=m,width=0.11\linewidth]{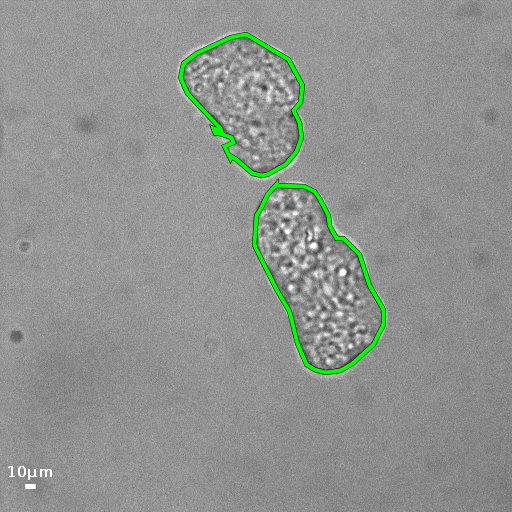}
		&\includegraphics[valign=m,width=0.11\linewidth]{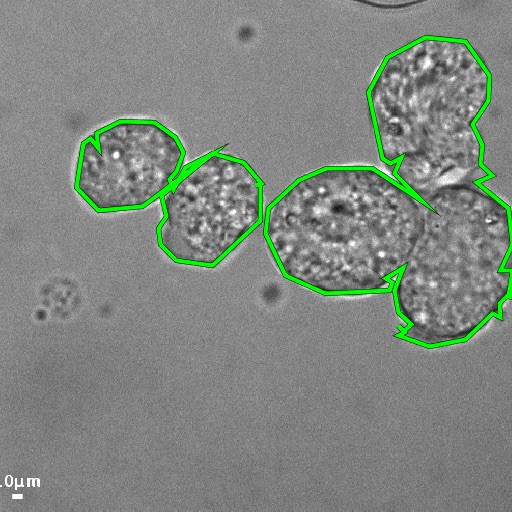}
		&\includegraphics[valign=m,width=0.11\linewidth]{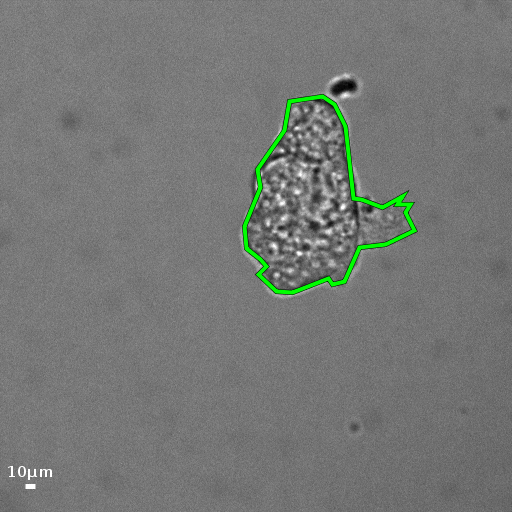}
		&\includegraphics[valign=m,width=0.11\linewidth]{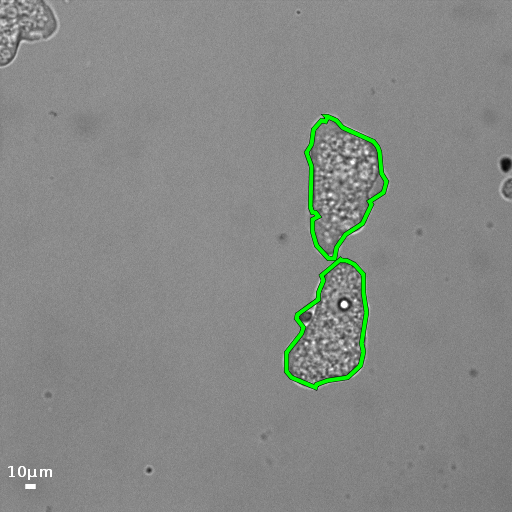}
		\\ 
		\scriptsize{\rotatebox[origin=c]{90}{ L2S}} &
		\includegraphics[valign=m,width=0.11\linewidth]{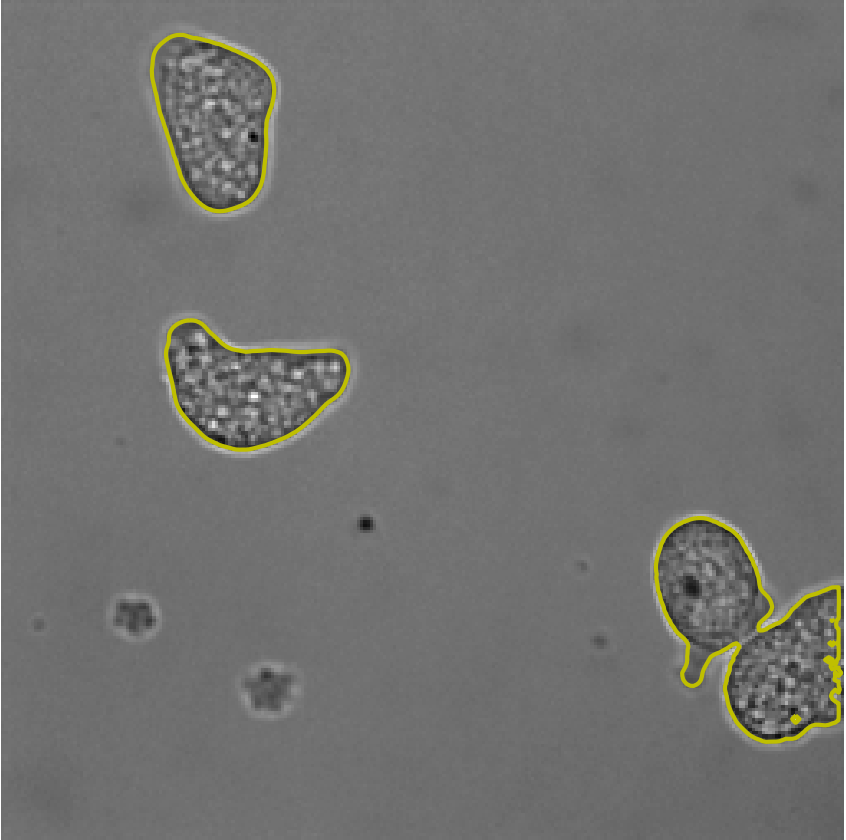}
		&\includegraphics[valign=m,width=0.11\linewidth]{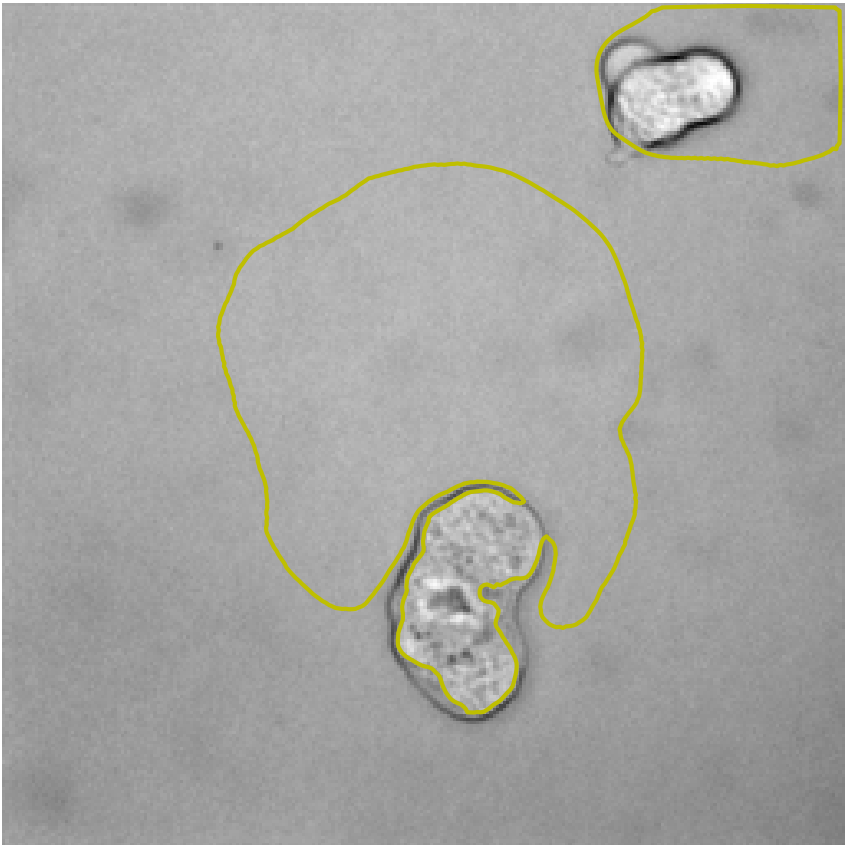}
		&\includegraphics[valign=m,width=0.11\linewidth]{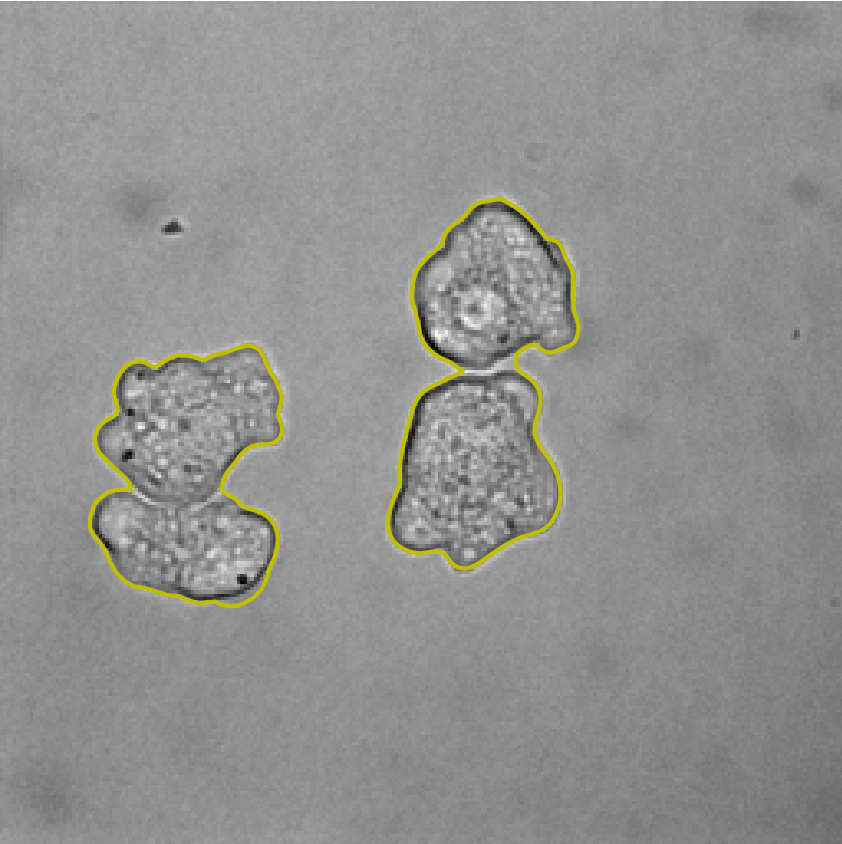}
		&\includegraphics[valign=m,width=0.11\linewidth]{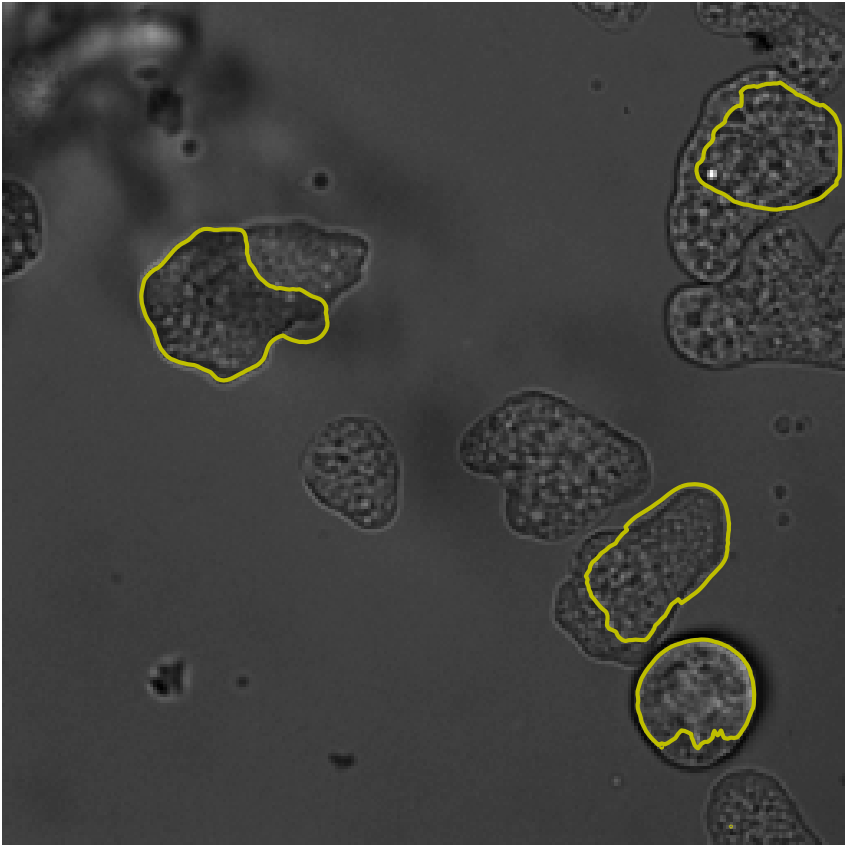}
		&\includegraphics[valign=m,width=0.11\linewidth]{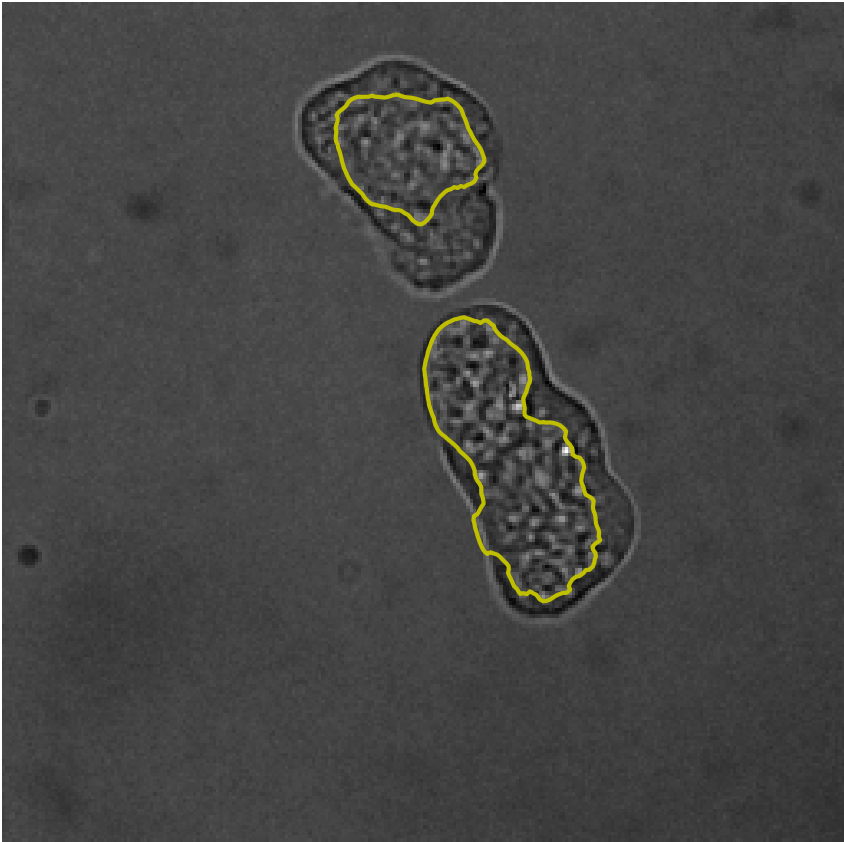}
		&\includegraphics[valign=m,width=0.11\linewidth]{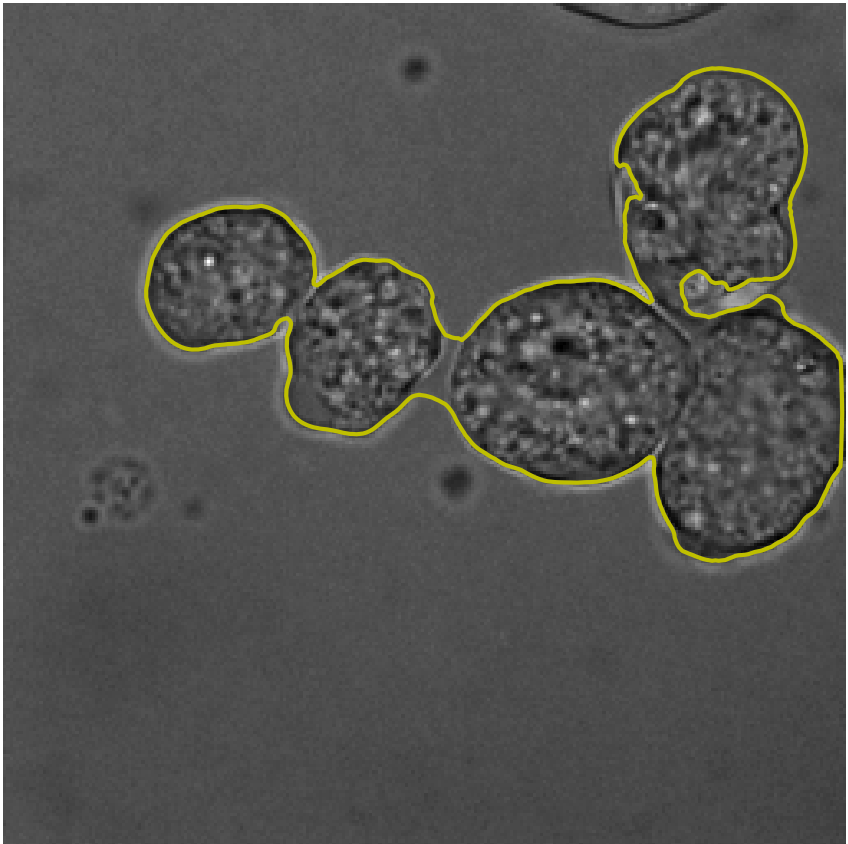}
		&\includegraphics[valign=m,width=0.11\linewidth]{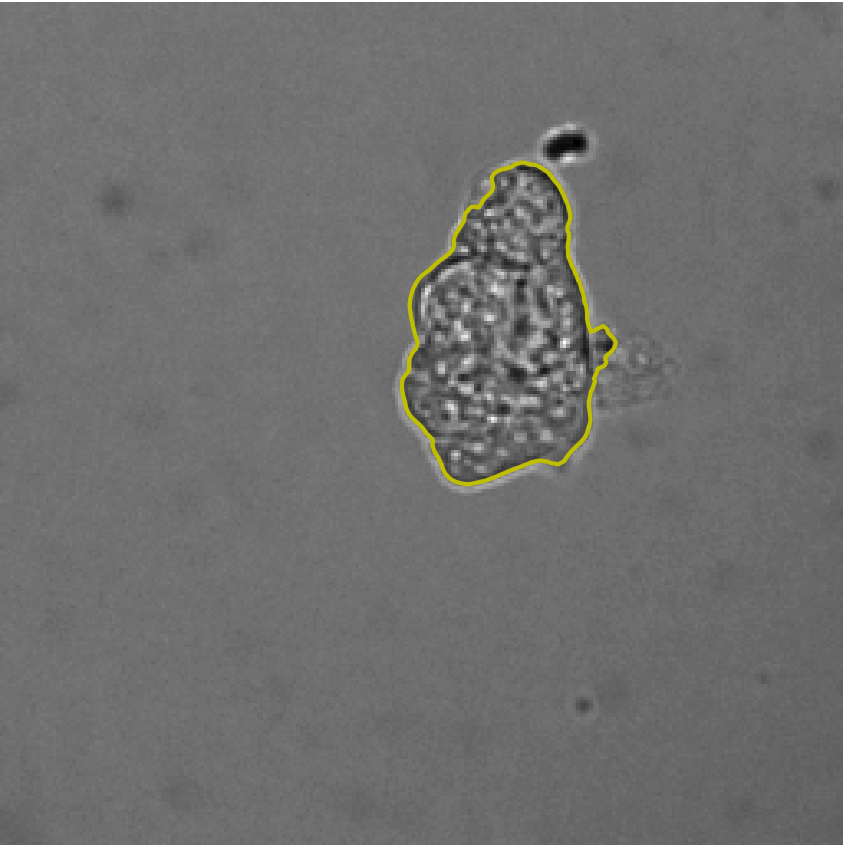}
		&\includegraphics[valign=m,width=0.11\linewidth]{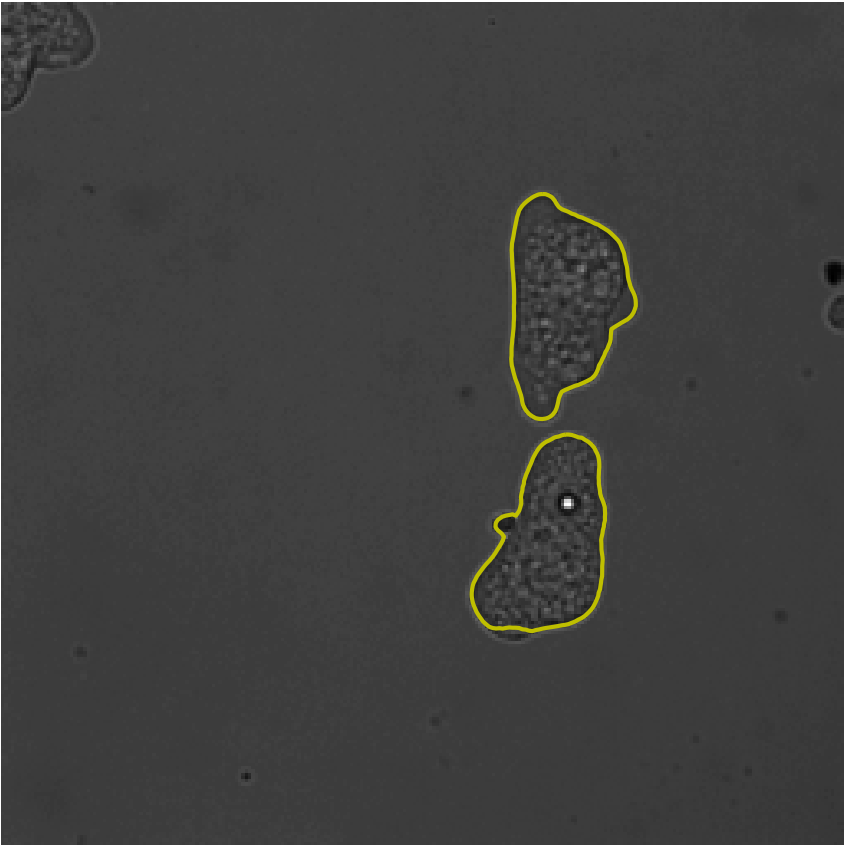}
		\\ 
		\scriptsize{\rotatebox[origin=c]{90}{Multi-CellNet}} &
		\includegraphics[valign=m,width=0.11\linewidth]{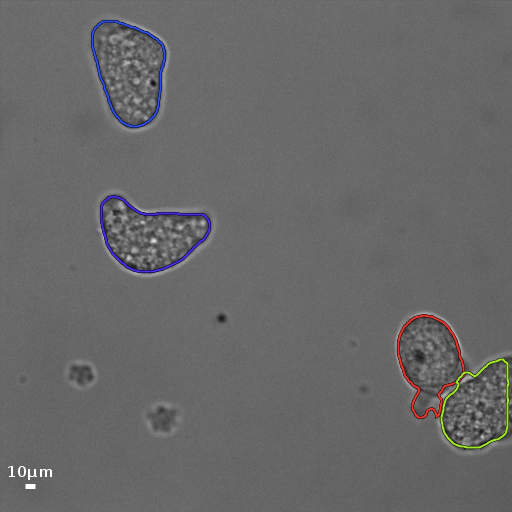}
		&\includegraphics[valign=m,width=0.11\linewidth]{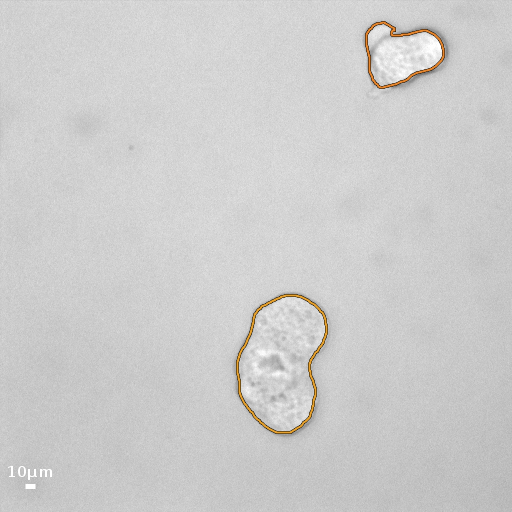}
		&\includegraphics[valign=m,width=0.11\linewidth]{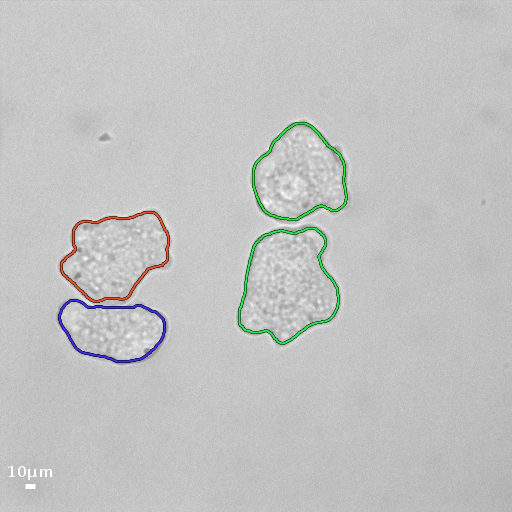}
		&\includegraphics[valign=m,width=0.11\linewidth]{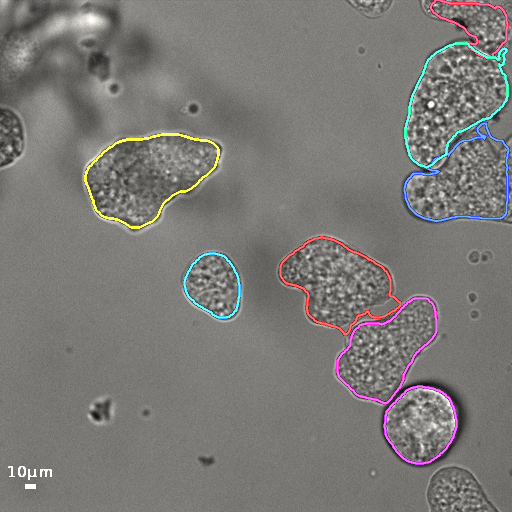}
		&\includegraphics[valign=m,width=0.11\linewidth]{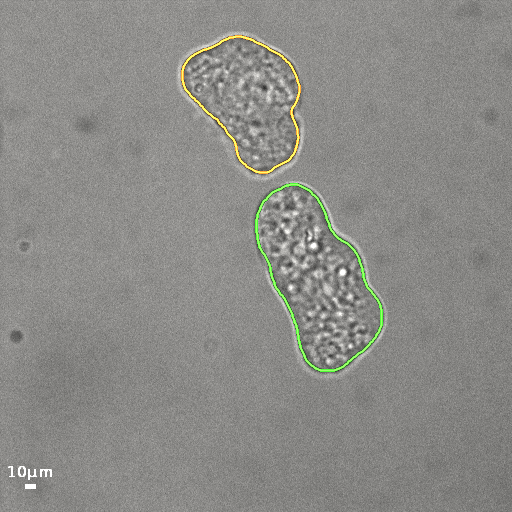}
		&\includegraphics[valign=m,width=0.11\linewidth]{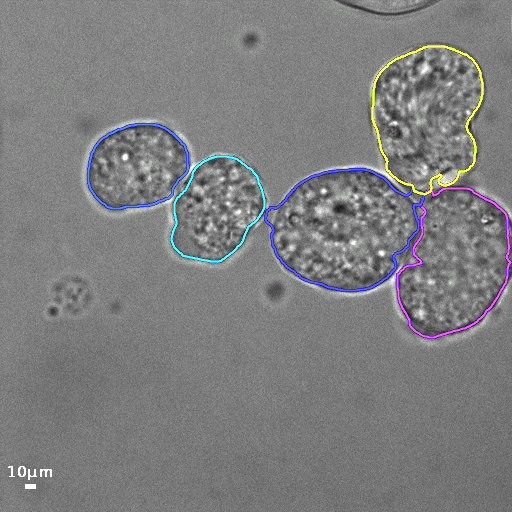}
		&\includegraphics[valign=m,width=0.11\linewidth]{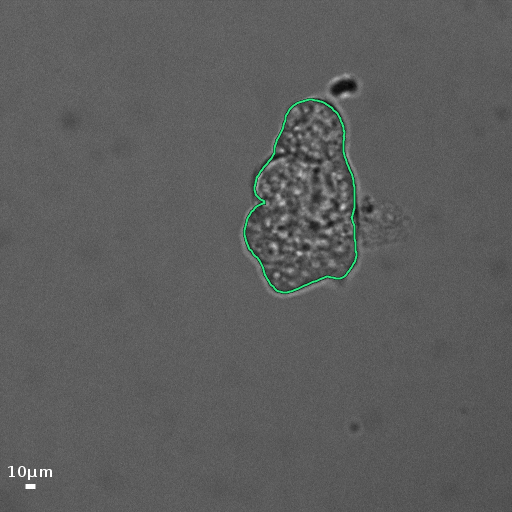}
		&\includegraphics[valign=m,width=0.11\linewidth]{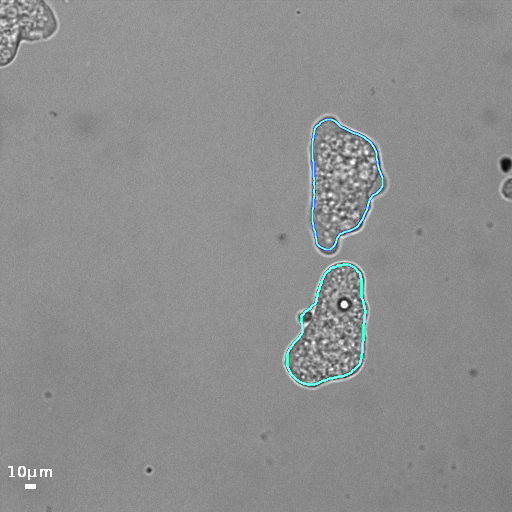}		
		\\
		\,&\small{(a)}&\small{(b)}&\small{(c)}&\small{(d)}&\small{(e)}&\small{(f)}&\small{(g)}&\small{(h)}
	\end{tabular}
	\caption[]{Eight representative brightfield images from our dataset are shown in the first row. The second and the third row shows the segmentation result due to U-net (green) \cite{ronneberger2015u} and L2S (yellow) \cite{mukherjee2014region}. The segmentation results of Multi-CellNet are shown in the bottom row, with the individual cell boundaries marked with different colors. The images are best viewed in color. The images have been contrast stretched for better viewing.}
	\label{fig: experimental results}
\end{figure*}

\textbf{Qualitative Evaluation:} The qualitative comparison of segmentation results are shown in Fig.~\ref{fig: experimental results}. The second row displays the object contours due to U-net in green color. Although U-net is quite suitable for segmenting separated cells, a few examples are shown in Fig.~\ref{fig: experimental results}(c), (d) and (f), where this method fails to identify the individual cells from a dense cluster. Specifically in scenarios  when the cells adhere along a considerable portion of their boundary, U-net is unable to detect separate cells from the cluster.
\begin{figure}[t]
	\centering
	\renewcommand{\tabcolsep}{0.05cm}
	\begin{tabular}{@{}ccc@{}}		
			\includegraphics[width=0.3\linewidth]{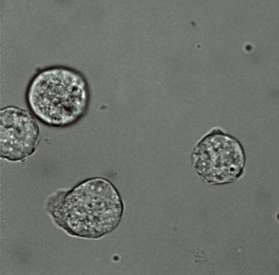}		
		&		\includegraphics[width=0.3\linewidth]{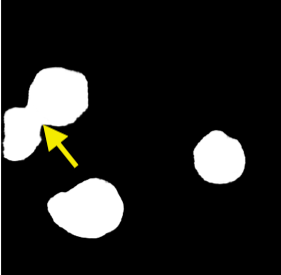}
				& \includegraphics[width=0.3\linewidth]{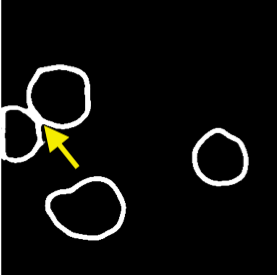}
				\\ \text(a) Original & \text{(b) Region} & \text{(c)\,Boundary}
			 \\
		\includegraphics[width=0.3\linewidth]{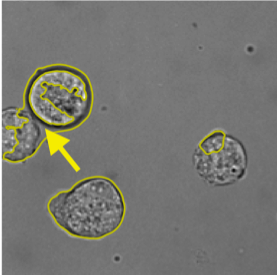}
		&\includegraphics[width=0.3\linewidth]{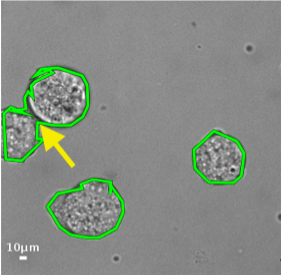}
		& \includegraphics[width=0.3\linewidth]{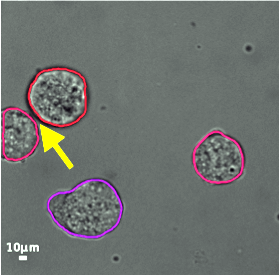}
			\\
		 \text{(d)\, L2S}&    \text{(e) U-net}	&\text{(f) \small{Multi-CellNet}}
		 
	\end{tabular}
	\caption[]{The figure exemplifies the effectiveness of Multi-AmoebaNet in delineating touching cells (marked by the arrow). The region and boundary detection from the Multi-AmoebaNet, segmentation result of L2s \cite{mukherjee2014region}, U-net \cite{ronneberger2015u}  and the the final segmentation after refinement of Multi-AmoebaNet is shown in (a), (b), (c), (d) and (e) respectively. Images best in color. Images have been contrast stretched for best viewing }
	\label{fig: qualitative_expl}
\end{figure}
Similarly, the segmentation accuracy of  L2S (shown in the third row in yellow color) is affected by cell density, and image contrast, and often fails to identify inter-cellular boundaries. Further, the effectiveness of L2S, as observed from experimentation, is dependent of initialization and image contrast. As mentioned, we employ $3^{rd}$ order Legendre polynomial for modeling the intensity in-homogeneity, which may not be the appropriate choice in for all the images and consequently the curve converges to a local minima.

The segmentation results due to Multi-CellNet are shown in the bottom row, and the individual cell boundaries are labeled with different colors. The qualitative results demonstrate the efficacy of Multi-CellNet to segment cells from dense clusters, especially in the presence of poor contrast and extensive shared boundaries between the cell walls (see Fig.~\ref{fig: experimental results} (d) \& (f)). A more challenging scenario is depicted in Fig. \ref{fig: qualitative_expl} (region marked in yellow).  
The effectiveness of Multi-CellNet over U-net and L2S is delineating touching cells can be noticed in more details in Fig. \ref{fig: qualitative_expl}. Although the separation between the touching cells is not detected by region detection (Fig. \ref{fig: qualitative_expl} (b)), the cell boundary is accurately detected by the edge-detection sub-network (Fig. \ref{fig: qualitative_expl} (c)). The appropriate consolidation of the two using the designed refinement step facilitates in isolating individual cells from a cluster (Fig. \ref{fig: qualitative_expl} (f)). 

\section{Conclusion}
\label{sec:conclusion}
In this paper, we propose a novel cell segmentation method leveraging the effectiveness of deep neural networks in a multi-task learning framework. Optimal performance of the multi-task learning network is achieved via adaptive combination of the loss functions. Final segmentation with smooth boundary is achieved using a dynamic clustering scheme of the region detection for curve initialization for an active contour model. The active contour is evolved using a speed function by appropriate consolidation of the region and edge predictions. Experimental validation demonstrates that the segmentation achieved with minimal refinement of the multi-task network predictions surpasses the state-of-the-art methods in various challenging scenarios. While the current model is effective for brightfield microscopy images only, future works will include experimentation and evaluation for fluorescence images of cells. 
\\
\textit{Acknowledgement}: This work was partially funded through grants from the Labex IBEID (ANR-10-LABX-62-IBEID), France-BioImaging infrastructure (ANR-10-INBS-04), and the programs PIA INCEPTION (ANR-16-CONV-0005) and DIM ELICIT Région Ile de France.

		 

\bibliographystyle{IEEEtran}	
\bibliography{ICPR_2020_v1}

\end{document}

%% file: zcom.tex
\newcommand{\beq}{\begin{equation}}
\newcommand{\eeq}{\end{equation}}
\newcommand{\bdm}{\begin{displaymath}}
\newcommand{\edm}{\end{displaymath}}

\newtheorem{definition}{Definition}

\newcommand{\bd}{\begin{definition}}
\newcommand{\ed}{\end{definition}}

\newcommand{\bv}{\begin{vugraph}}
\newcommand{\ev}{\end{vugraph}}
\newcommand{\bi}{\begin{itemize}}
\newcommand{\ei}{\end{itemize}}
\newcommand{\ben}{\begin{enumerate}}
\newcommand{\een}{\end{enumerate}}

\newcommand{\bean}{\begin{eqnarray*} }
\newcommand{\eean}{\end{eqnarray*} }
\newcommand{\bea}{\begin{eqnarray} }
\newcommand{\eea}{\end{eqnarray} }

\newcommand{\ba}{\begin{array} }
\newcommand{\ea}{\end{array} }

